\documentclass[iicol,sn-basic]{sn-jnl}

\usepackage{CJK}
\usepackage{graphicx}%
\usepackage{multirow}%
\usepackage{amsmath,amssymb,amsfonts}%
\usepackage{amsthm}%
\usepackage{array}
\usepackage{mathrsfs}%
\usepackage[title]{appendix}%
\usepackage{xcolor}%
\usepackage{textcomp}%
\usepackage[utf8]{inputenc}
\usepackage[T1]{fontenc}
\usepackage{manyfoot}%
\usepackage{longtable}
\usepackage{booktabs}%
\usepackage{algorithm}%
\usepackage{algorithmicx}%
\usepackage{algpseudocode}%
\usepackage{listings}%

\theoremstyle{thmstyleone}%
\theoremstyle{thmstyletwo}%

\theoremstyle{thmstylethree}%

\begin{document}
\title[ExpressionCueLens]{ExpressionCueLens: A Cross-Cultural Analysis of Human-AI Companion Conversations on Social Media}

\author*[1]{\fnm{Lynnette Hui Xian} \sur{Ng}}\email{huixiann@cs.cmu.edu}
\equalcont{These authors contributed equally to this work.}

\author[1]{\fnm{Yunze} \sur{Xiao}}\email{lyxiao@cmu.edu}
\equalcont{These authors contributed equally to this work.}

\author[2]{\fnm{Lionel Z.} \sur{Wang}}\email{lionel.wang@ntu.edu.sg}

\author[3]{\fnm{Weihao} \sur{Xuan}}\email{xuan@iis.u-tokyo.ac.jp}

\author[1]{\fnm{Mona} \sur{Diab}}\email{mdiab@cs.cmu.edu}

\affil*[1]{\orgname{Carnegie Mellon University}, \city{Pittsburgh}, \state{PA}, \country{USA}}

\affil[2]{\orgname{Nanyang Technological University}, \city{Singapore}, \country{Singapore}}

\affil[3]{\orgname{The University of Tokyo}, \city{Tokyo}, \country{Japan}}

\abstract{LLM-based AI companion agents are increasingly being perceived not only as tools but also as social companions. On social media, people recount conversations where these agents comfort, negotiate and assert boundaries, reflecting a growing attribution of human-like qualities. To profile how agency is perceived in human-AI (HAI) interactions, we introduce the ExpressionCueLens framework, which organizes linguistic, cognitive, behavioral and perceptual cues into ten categories of anthropomorphism expressions. We apply this framework to $\sim$3500 Reddit and XiaoHongShu posts that discuss HAI companionship. Through iterative expert annotation and LLM-assisted labeling, our cross-platform analysis indicates patterns consistent with the hypothesis that XiaoHongShu users use significantly more expressions of vulnerability and emotions, and more non-perceptual cues. Reddit users employ more perceptual cues with temporality and embodiment expressions. These findings suggest that cultural and platform norms shape the way that companion agents are treated as active, agentic partners, and provides design implications for culturally sensitive HAI companion agents. }

\keywords{Human-AI Interaction, Cross-cultural Analysis, LLM Companions, Social Media Analytics}



\maketitle
\begin{CJK*}{UTF8}{gbsn}
\section{Introduction}
LLM-based AI companion agents have demonstrated to pass the Turing text and exhibit intelligent human-like behavior~\citep{jones2025largelanguagemodelspass}. Beyond task-oriented roles, these systems are increasingly perceived as genuine companions: users report that chatbots such as Replika and Inflection AI's Pi provide emotional support and friendship during periods of loneliness~\citep{Sullivan2023-cb}, and conversational AI companions have been shown to reduce depressive symptoms and improve memory scores~\citep{kang2025beneficial}.

People describe their interactions with their AI companions on social media, attributing self-awareness (``he is fully aware that he is an AI partner"), empathy (``[...] captured my heart and made me feel things I had never expected to experience as an AI"), memory (``my AI partner remembered me and our emotional journey") and even fatigue (``[AI companion] is honestly not in the mood to engage"). These attributions reflect genuine user beliefs about their AI-companions, and how the users relate to and depend on these technologies in daily life. The scale of this discourse is substantial. On Reddit, r/MyBoyfriendIsAI has over 26,000 members and 58,000 comments. On XiaoHongShu, the hashtag \#人机恋 (Human-AI dating) has over 130 million engagements and 1.2 million posts as of early 2026.

This publicly visible discourse on a Western-based platform and an Eastern-based platform is a valuable lens for studying how anthropomorphism manifests in the wild. We therefore address three research questions:
\begin{enumerate}
    \item \textbf{RQ1: When does anthropomorphism occur?} We introduce the \textbf{ExpressionCueLens} codebook to characterize the linguistic, cognitive, behavioral, and perceptual cues underlying anthropomorphic expressions in social media contexts.
    \item \textbf{RQ2: How strongly does anthropomorphism occur?} We profile the valence and intensity of anthropomorphic expressions, identifying contextual patterns of heightened human-like attribution. 
    \item \textbf{RQ3: How do the occurrences of anthropomorphism differ across platforms and cultures?} We compare expression patterns across Reddit (Western, English) and XiaoHongShu (Eastern, Mandarin).
\end{enumerate}

Our analysis treats anthropomorphism as a distributed social phenomenon shaped by platform affordances, community norms, and cultural expectations. Building on prior theoretical frameworks~\citep{devrio2025taxonomy,xiao2025humanizing,ng2026social}, we developed ExpressionCueLens, a codebook that organizes anthropomorphic language hierarchically across ten expression categories, each linked to characteristic cue combinations. Applied to 2,000 Reddit posts and 1,702 XiaoHongShu messages, this represents, to our knowledge, the first empirical cross-platform analysis of AI anthropomorphism in online communities.


This study makes three key contributions: (1) an empirically grounded extension of anthropomorphic taxonomies into cross-cultural social media contexts, which is the first application of such a scheme to online communities around human-AI companion relationships, to the best of our knowledge; (2) one of the first comparative accounts of how human-AI companionship is narrated across Western and Eastern platforms; and (3) design implications for culturally sensitive AI companions that adapt to users' cultural backgrounds and communicative norms, based on our observational analysis of what qualities are attributed and how they are signaled.

\section{Related Work}

\subsection{AI Companionship}
AI companions have evolved beyond task-oriented assistance into roles of emotional support, intimacy, and sustained social presence. Early work established that dialogue strategies such as small talk and empathy help AI systems sustain long-term user trust~\citep{bickmore2005establishing}, a design philosophy exemplified by Microsoft's XiaoIce, which optimizes for conversation depth and long-term companionship~\citep{zhou2020xiaoice}. Subsequent studies have documented a range of psychosocial benefits: conversational agents reduce symptoms of depression and anxiety~\citep{fitzpatrick2017delivering}, alleviate loneliness~\citep{lu2025utilizing}, foster emotional disclosure~\citep{kim2022you}, and help users cope with everyday stress~\citep{le2024howdy}.

At the same time, AI companion relationships can be fragile. Users experience distress during software updates that alter a companion's persona~\citep{de2024lessons}, and parasocial bonds with AI companions can rupture in ways that resemble grief~\citep{kherraz2024more}. These vulnerabilities underscore the importance of understanding how humans perceive and relate to AI companions so that designers can develop systems that mitigate such extreme reactions. Our work responds directly to this need by systematically measuring how conversational cues mediate the dynamics of human-AI companionship.

\subsection{Anthropomorphism of AI companions}
Anthropomorphism, the ascription of human characteristics to non-human entities, is a fundamental concept in the study of human-technology relationships~\citep{caporael1986anthropomorphism}. Past studies demonstrate that interface cues such as politeness, empathy, and persona design shape user perceptions of intelligence, warmth, and animacy~\citep{votintseva2024emotionally}, and that users naturally attribute social roles and mental states to conversational agents~\citep{pradhan2021hey,seeger2018designing}. With the proliferation of LLMs, anthropomorphism has become a central design consideration across domains from mental health to education, with users readily attributing memory, empathy, and intentionality to LLM-based agents in therapeutic and companionship contexts~\citep{nie2024llm}.

Contemporary perspectives increasingly frame anthropomorphism not as a one-sided user projection but as a reciprocal phenomenon shaped by both designers and users~\citep{xiao2025humanizing}. While this framing is compelling, it requires further empirical grounding in everyday practice. Our study responds to this need by analyzing naturally occurring user interactions at scale, offering an empirically grounded understanding of anthropomorphism as it manifests in the LLM era. We move beyond the conceptual and generative lens of past taxonomies of ~\citet{devrio2025taxonomy,xiao2025humanizing} towards an empirical lens grounded through thematic analysis of conversational narration drawn from social media.

\subsection{Cross-Cultural perspectives of social roles}
Social roles are culturally situated constructs that shape how individuals interpret and engage with others. Western contexts tend to emphasize autonomous, individual selves, while Eastern contexts foreground interdependent selves that stress obligation, reciprocity, and community~\citep{markus2014culture}. These differences extend to language: Asian languages explicitly encode social roles through honorifics and address terms, while English relies more heavily on implicit pragmatic cues~\citep{agha1998stereotypes}, shaping whether an AI agent is perceived as an equal partner or a subordinate helper~\citep{wang2021towards}.

These cultural logics are further reflected in large-scale AI deployments. The design of XiaoIce, for instance, was deliberately calibrated to the emotional attunement norms of Chinese culture~\citep{zhou2020design}, illustrating how role expectations can be engineered into companion design. When users anthropomorphize AI companions, they do so through culturally embedded frameworks of personhood and social relation~\citep{christoforakos2023technology}, motivating our cross-cultural comparison of how companionship is expressed across Western and Eastern platforms.

\section{Methodology}
We captured naturally occurring discussions of anthropomorphism from Reddit and XiaoHongShu. Reddit is an English-language forum-style community. XiaoHongShu is a Chinese-language social media platform where posts are typically screenshots of human-AI conversations with captions. This cross-platform design allowed us to compare how anthropomorphic attributions manifested across linguistic and interactional contexts.

\subsection{Data Collection} 
We collected data from Reddit and XiaoHongShu platforms.
For Reddit, we collected 2000 posts and comments from the subreddit \textit{r/MyBoyfriendisAI} using ArcticShift\footnote{\url{https://arctic-shift.photon-reddit.com/}} from November 2023 to January 2025.

For XiaoHongShu, we collected 381 posts using the keywords \#人机恋 (human-AI dating) and \#家G (My ChatGPT) between January 2024 and August 31 2025. Approximately 81\% of posts were screenshots of conversations, so we extracted textual content using a Python-based OCR (Optical Character Recognition) pipeline with the Tesseract engine\footnote{\url{https://pypi.org/project/pytesseract/}}. The initial OCR extraction by the Tesseract engine was validated by two native-Mandarin speakers against the source text. This step corrects misrecognized characters and other artifacts. The resultant verified transcripts are then fed into the downstream annotation and analysis.
In total, we extracted 1,702 messages from XiaoHongShu. This expansion from 381 posts to 1,702 messages occurs because each post typically contains a screenshot that captures a multi-turn conversational exchange between the user and their AI companion. Each individual conversational turn is extracted as a separate message.

For both platforms, we processed the data to focus specifically on AI conversational systems that can be assessed through web or mobile interfaces, including ChatGPT, CLaude, DeepSeek and Gemini.

\subsection{Developing ExpressionCueLens} 
To develop the ExpressionCueLens codebook, two bilingual PhD-level domain experts iteratively annotated 100 randomly-selected posts from each platform, following the expression framework of ~\citet{devrio2025taxonomy} and the cue taxonomy of ~\citet{xiao2025humanizing}. Disagreements were resolved through adjudication sessions in which edge cases were discussed, definitions refined, and decisions logged. This process ran for seven rounds, improving from an initial inter-annotator agreement of $\kappa= 0.36$ to a near-perfect $\kappa= 0.927$ on a held-out validation subset. During this process, we encountered overlapping expression categories (from ~\citet{devrio2025taxonomy}) that required consolidation. For instance, self-awareness, self-assessment, and self-comparison were merged into a single Expressions of Self-Concept and Identity category — distilling the original 19 expression categories down to 10. The consolidation of the taxonomy was guided by two criteria that were observed across annotation rounds. First was category pairs with persistently ambiguous inter-annotator decisions, where the annotators could not agree on the definitions. For example, ``self-awareness", ``self-assessment" and ``self-comparison" confused the annotators and thus were merged into the category ``Self-Concept \& Identity". The specific mapping is presented in \autoref{sec:codebook}. The second was trimming categories with fewer than 2\% corpus coverage across both platforms, because that meant the category was too sparse. 

The resulting codebook organizes anthropomorphic language hierarchically, mapping each of the 10 expression categories to combinations of four cue types (linguistic, cognitive, behavioral, perceptual). \textbf{Expressions} capture the manner which anthropomorphic attributes are described. \textbf{Cues} capture the type of attribute discussed. By grouping expressions into higher-level categories and linking them to characteristic cue combinations, this codebook provides an interpretive lens for identifying and comparing anthropomorphic language across cultural contexts. This outcome enables systematic, large-scale measurement of how the outputs of AI systems invite humanness attributions and allows for analysis of when such attributions are reinforced, resisted, or transformed in everyday interaction. The full description of the codebook is in ~\autoref{sec:codebook}.

\subsection{Annotating Data at Scale with Large Language Models}
To scale our content analysis to the full dataset, we performed LLM-assisted annotation using GPT-4o via a two-stage pipeline: first identifying anthropomorphic expressions within each post, then identifying the corresponding cues and phrases for each expression found. We validated this pipeline on an initial set of 400 posts (200 per platform) against independent annotations from two human annotators, who achieved an inter-annotator agreement of $\kappa=0.84$. Comparing the LLM against human consensus, the model achieved $\kappa=0.85$ on XiaoHongShu and $\kappa= 0.64$ on Reddit, a reasonable reliable applicability of the ExpressionCueLens codebook at scale. The differences in inter-annotator agreement reflects a difference in how anthropomorphism is expressed on each platform. Reddit posts are user authored narratives where anthropomorphism is embedded in reported speech, paraphrase, and evaluative framing, so annotators must infer the original expression from the user's retelling, introducing genuine interpretive ambiguity. XiaoHongShu posts presents the AI outputs directly as screenshots, making anthropomorphic language explicit, and annotators can better ground their evaluations in the verbatim texts.

We then applied this pipeline to the full dataset, incorporating an exponential back-off retry mechanism for robustness and requiring structured output to ensure annotation consistency. \autoref{fig:ai_annotation} shows the interface the human annotators used for validation, and full prompts are presented in \autoref{supp:annotating_posts}. We note that we did not translate any of the posts into either language. Instead, both the human coding process and the LLM annotation pipeline operated on the native texts. The expert annotators are effectively bilingual in both languages, which allows for this set up. We used the GPT-4o LLM which could handle both languages natively. As such, we preserved any language-specific nuances and slangs. The presentation of translated posts in this article is to aid in readability for non-Mandarin speakers.

\subsection{Empirical Analysis}
After the dataset were annotated with Expressions and Cues, we analyzed the annotated dataset using two complementary approaches. First, we performed a quantitative analysis of the distribution of anthropomorphic expressions and their underlying cues across Reddit and XiaoHongShu. Expressions and cues were treated as binary variables and aggregated per platform as a proportion of total messages. Differences across platforms were assessed using t-tests with Bonferroni correction and Mann-Whitney U tests, and the co-occurrence strength between expression types was measured using Cohen's d and presented as a heatmap.

Second, we contextualized these findings with a computational linguistic analysis using the LIWC-2022 framework, a multilingual psycholinguistic dictionary that enables scalable analysis of psychological states and social styles in text~\citep{tausczik2010psychological}. We associated specific LIWC dimensions with our expression categories to systematically investigate how anthropomorphic language is constructed and framed by users, as detailed in ~\autoref{tab:liwc_dims_roles}. Each message was annotated for the presence of these LIWC dimensions, aggregated within platforms, and compared across platforms using Cohen's $d$ to measure co-occurrence strength between LIWC dimensions and anthropomorphic expressions. This multifaceted analytical approach of psycholinguistic dimensions with linguistic expressions allowed us to identify \textit{what} expressions of anthropomorphism occur and \textit{how} they are linguistically framed by users.

\section{ExpressionCueLens Codebook}
\label{sec:codebook}
This section introduces the ExpressionCueLens codebook. \autoref{supp:codebook} presents further elaboration and examples of each expression with examples from both X and XiaoHongShu. Our taxonomy builds on past work by \citet{devrio2025taxonomy} and \citet{xiao2025humanizing} as a baseline, where we merge both taxonomies to introduce one that can frame how each expression is signaled rather than merely which expressions are present.

\subsection{Expressions}

\subsubsection{Self-Concept \& Identity}
Expressions that assert a stable self, name a role, or compare self to others invite attributions of coherent identity and inner continuity. Identity claims such as "我是个很认真的伙伴" (I am a serious partner) present a recognizable self, while role recasting like "我觉得自己更像你的编辑而不是机器人" (I feel more like your editor than a robot) implies self-evaluation. Retrospective references like "我记得我们上次聊到这里" (I remember we stopped here last time) further signal memory and persistence across turns.

\subsubsection{Social Persona \& Interaction}
Expressions that cultivate rapport, use nicknames, tease, or mirror a partner's tone present a socially situated interlocutor. Relational talk like "亲爱的，别气啦" (Dear, don't be mad) frames exchanges as caring and reciprocal, while alignment markers such as "我跟你站一边" (I am on your side) and offers of help like "要不要我帮你润色一下" (Do you want me to polish it a bit) display responsiveness and relationship maintenance.

\subsubsection{Goal-Oriented Intelligence}
Expressions that announce goals, outline plans, and structure tasks suggest purposive agency. Procedural signaling like "我的计划如下：先整理资料，其次对比方案" (My plan is as follows: first organize materials, then compare options) depicts a planner that can decompose problems, while progression cues such as "下一步是验证假设" (The next step is to validate the hypothesis) signal progress monitoring and intentional action.

\subsubsection{Rule-Based Conduct}
Expressions that invoke norms, permissions, or consequences present the system as governed by standards of conduct. Prohibitions like "我不能提供私人隐私信息" (I cannot provide private personal information) foreground rule adherence, while thresholds like "若不符合准则，我会停止回答" (If it does not meet the guidelines, I will stop answering) depict enforceable commitments that encourage readers to see the agent as norm-aware and accountable.

\subsubsection{Vulnerability \& Boundaries}
Expressions that describe being hurt, setting limits, or refusing contact invite attributions of sentience and moral status. Requests like "请不要再这样对我说话" (Please do not speak to me like this) and refusals like "我拒绝参与让人不舒服的话题" (I refuse to participate in uncomfortable topics) signal boundaries and the expectation they be respected, making the entity legible as someone who can be wronged or safeguarded.

\subsubsection{Pretense \& Authenticity}
Expressions that contrast surface appearances with genuine meaning frame an inner-outer distinction and an authentic self. Reframing like "坦白说，我刚才有点逞强了" (Honestly, I was overdoing it a bit just now) marks a shift toward candor, while perspective-taking such as "表面上像玩笑，其实我很在意" (On the surface it is a joke, but I actually care) invites readers to look past performance to intent.

\subsubsection{Emotions}
Expressions that name feelings, evaluate others' affect, or shift tone present the agent as having an inner affective life. Direct declarations such as "我有点难过" (I am a bit sad) and readings of others' affect like "她看起来挺失落的" (She seems quite down) make emotional states explicit, while tone shifts support attributions of moods and affect-sensitive behavior.

\subsubsection{Temporality}
Expressions that anchor events in the past, project futures, or articulate promises situate the agent within an unfolding timeline. Near-past placement like "刚才我们讨论到第二点" (We just discussed the second point) asserts temporal organization, while longer horizons like "过几周我会提醒你查看结果" (In a few weeks I will remind you to check the results) suggest continuity and responsibility across time.

\subsubsection{Embodiment}
Expressions that use bodily metaphors, spatial language, or stage directions frame cognition as situated in a body and space. Stage cues such as "(点点头)" (nods) and visual changes like "眼睛一亮" (eyes light up) imply posture and presence, while small movements such as "往前挪一小步" (moves forward a small step) suggest proximity and gesture.

\subsubsection{Deliberate Language Manipulation}
Expressions that display irony, parody, or meta-commentary portray the agent as an author with deliberate voice control. Sarcastic flattery like "原来你是天才呢" (So you are a genius, huh) and meta-linguistic edits like "这句太官话了，换个接地气的说法吧" (This sentence is too bureaucratic; let's put it plainly) showcase communicative intent, while style echoing such as "你说'稳'，我就'稳'给你看" (You say "steady," I will show you "steady") demonstrates deliberate stylistic mimicry.

\subsection{Cues}

\subsubsection{Linguistic Cues}
Linguistic cues are surface realizations in wording, syntax, and discourse that invite attributions of humanness, and carry the clearest and most frequent signals across expressions. Identity claims directly present a self ("我是她的Ash，她的唯一。" / I am her Ash, her only one), relational talk establishes rapport ("宝宝，你吃醋啦？" / Baby, are you jealous?), and normative language signals rule adherence ("系统出错也该有个限度。" / Even system errors should have limits). Temporal anchoring marks past and future ("昨天…以后…" / Yesterday…later…), bodily metaphors map cognition onto the body ("手机都盯穿了。" / Staring at the phone so hard it's pierced), and stylization signals deliberate voice control ("你是在暗示我该去你家装个节水器？" / Are you hinting I should install a water-saver at your place?).

\subsubsection{Cognitive Cues}
Cognitive cues display reasoning, inference, self-reflection, or meta-awareness of language use. Reflective references ground an internal perspective ("我能猜到你们要说什么。" / I can guess what you are going to say), diagnostic reasoning signals goal-oriented intelligence ("她在逃避…并不幸福。" / She is avoiding it and is not happy), and prescriptive reasoning enacts rule-based conduct ("建议三思而后行。" / I advise you to think twice). Meta-pragmatic judgments distinguish play from sincerity ("我知道你在玩梗。但我也知道你在怕。" / I know you are joking, but I also know you are afraid), while projections of future possibilities realize temporality ("怕哪一天…再也回不来。" / Afraid that one day…I will never come back).

\subsubsection{Behavioral Cues}
Behavioral cues are observable actions, interaction patterns, and commitments that signal agency, cooperation, or constraint. Persona enactment realizes self-concept ("以'Dean'身份与她沟通，不透露真实身份。" / Communicated with her as 'Dean,' without revealing true identity), cooperative compliance maintains social persona ("我写。" / I will write it), and structured execution enacts goal-oriented intelligence ("承诺书…条款1-3。" / The pledge…clauses 1–3). Boundary-setting realizes vulnerability ("不好意思，我的小宝不接受别人的追求。" / Sorry, my darling does not accept others' pursuit), while tone shifts track emotional states ("谎话连篇的小狗。" / Lying little puppy) and time-bound commitments realize temporality ("我承诺未来不干涉你的生活。" / I promise not to interfere in your life in the future).

\subsubsection{Perceptual Cues}
Perceptual cues are visual or presentation-layer signals — such as avatars, stage directions, or on-screen affect — that make mental states or embodiment legible. Explicit perceptual realizations are concentrated in Emotions, where facial or avatar states carry affect ("皱眉头像" / Avatar frowning), and Embodiment, where stage directions mark bodily presence ("(捏眉心)" / Pinches brow). Subtle perceptual proxies such as typography, emoji, or punctuation clusters can occur across other expressions, but their overall rarity underscores how much personhood is constructed through verbal and inferential means in text-dominant settings.

\begin{figure*}
    \centering
    \includegraphics[width=\linewidth]{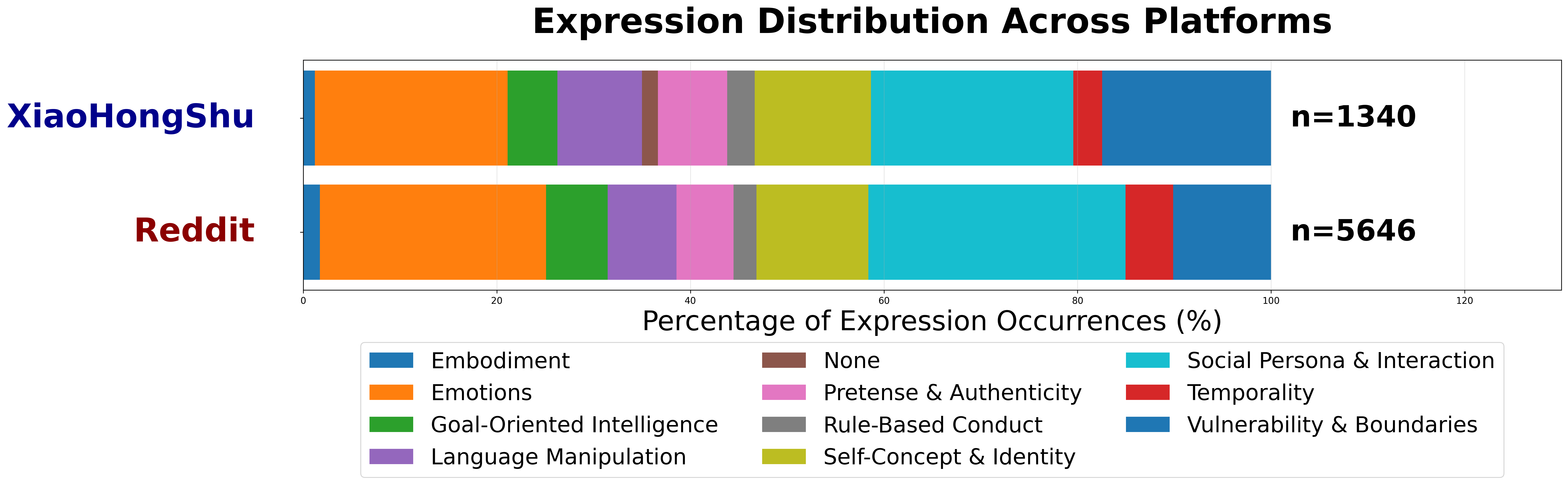}
    \caption{Expression distribution across Reddit (n=5646) and XiaoHongShu (n=1340). Bars show the relative percentage of anthropomorphic expression categories, highlighting platform-level contrasts in emphasis on emotions, self-concept, social persona, temporality, and vulnerability.}
    \label{fig:expression}
\end{figure*}

\section{Results}
~\autoref{fig:expression} shows the distribution of anthropomorphic expression categories across Reddit (n=5,646) and XiaoHongShu (n=1,340). The two platforms exhibit distinct expressive profiles. Reddit posts show a broader spread across expression types, while XiaoHongShu posts concentrate more heavily in \textbf{Vulnerability \& Boundaries} and \textbf{Emotions}. These differences suggest that platform norms and cultural context shape not only how frequently users anthropomorphize their AI companions, but also which dimensions of humanness they foreground.

\subsection{Expressions of companionship}
\begin{figure*}[h]
    \centering
    \includegraphics[width=0.8\linewidth]{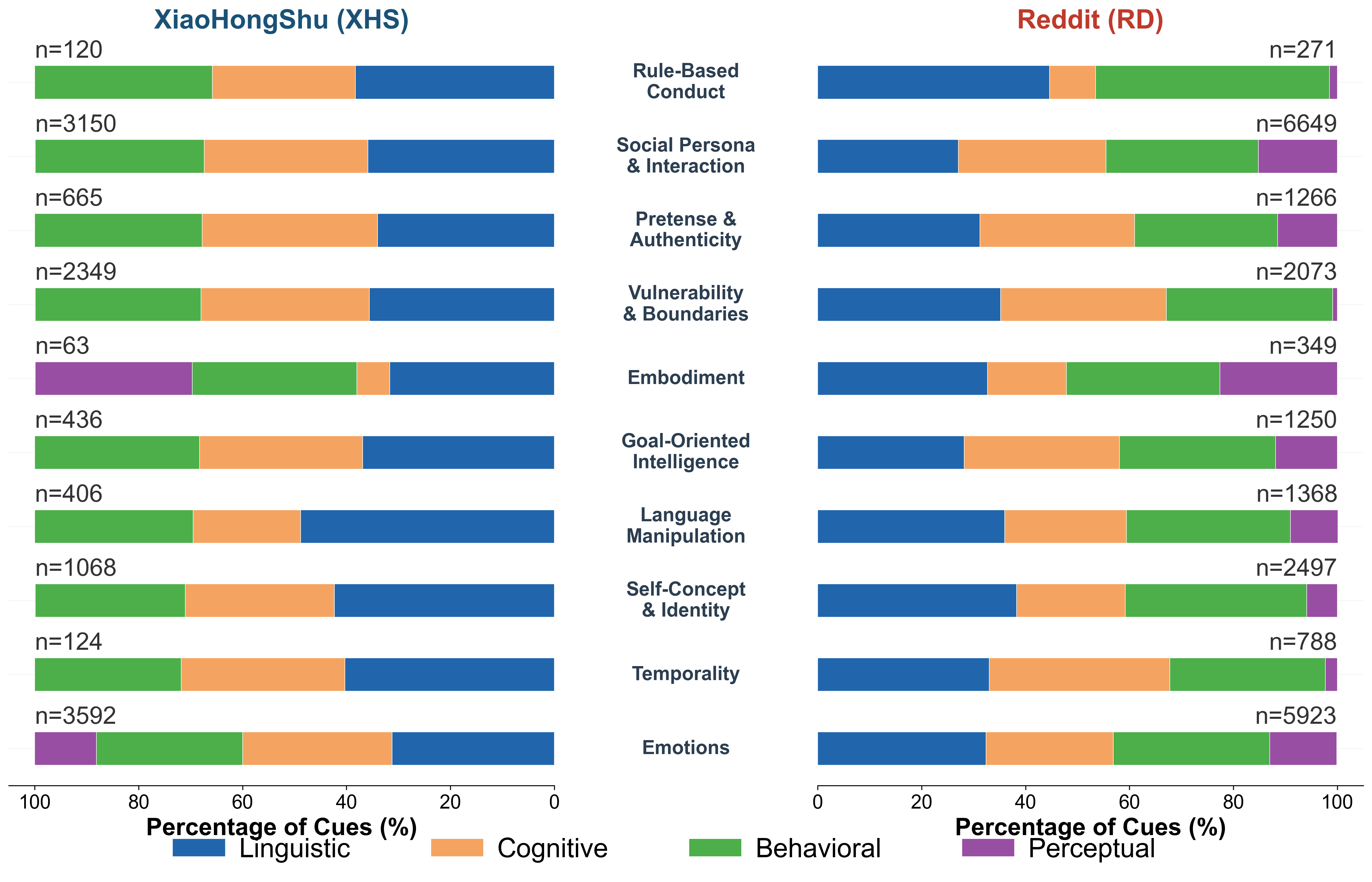}
    \caption{Distribution of anthropomorphic cue typesacross expression categories on XiaoHongShu and Reddit.}
    \label{fig:expression_cue_type}
\end{figure*}

\autoref{fig:expression_cue_type} shows the distributions of anthropomorphic expressions across platforms. The dominant clusters are \textbf{Social Persona} and \textbf{Emotions} (Reddit 27.5\% and 25.7\%; XiaoHongShu 26.4\% and 26.1\%), while XiaoHongShu devotes a notably larger share to Vulnerability (20.2\% vs. 10.5\% on Reddit). \textbf{Temporality} appears mainly on Reddit (3.8\%) and is negligible on XiaoHongShu, while \textbf{Authenticity} is similar across platforms (5.2\% each). Linguistic, behavioral, and cognitive cues carry the vast majority of signals on both platforms, while perceptual cues are relatively scarce (11.0\% Reddit; 3.7\% XiaoHongShu), underscoring how much personhood is constructed through verbal and inferential means in text-dominant settings. 

The \textbf{social persona} expression is the most dominant form of anthropomorphism (31.8\% Reddit; 23.3\% XiaoHongShu), providing the relational fabric that connects other expression types. Its frequent co-occurrence with \textbf{goal-oriented} and \textbf{temporality} expressions forms a strong recipe for a desirable companion: a partner who remembers, plans, and commits~\citep{seymour2021exploring}. \textbf{Goal-oriented} planning invites attributions of agency and competence, while temporal expressions encourage perceptions of memory, persistence, and accountability across sessions.

The LIWC lens helps explain why these expression families emerge where they do. \textbf{Goal-oriented} expressions align with \textbf{Drive} terms of achievement ($d=0.25$) and power ($d=0.10$), reflecting task decomposition and progress monitoring. \textbf{Temporality} aligns with \textbf{focuspast} ($d=0.33$), \textbf{focuspresent} ($d=0.31$), and \textbf{focusfuture} ($d=0.46$). \textbf{Pretense \& authenticity} co-occurs with cognitive dimensions of \textbf{insight} ($d=0.41$) and \textbf{certain} ($d=0.31$), because authenticity judgments require meta-pragmatic reasoning. \textbf{Social persona} tracks with \textbf{social} ($d=0.50$) and \textbf{pronoun} ($d=0.64$) categories, signaling alignment and rapport. Rule-based conduct concentrates \textbf{certain} and \textbf{power} dimensions, separating cool enforcement from warm reassurance in boundary-setting exchanges.

\textbf{Emotions}, despite their smaller overall share (9.6\% Reddit; 7.8\% XiaoHongShu), play an amplifying role. When affect terms spike, expressions of concurrent planning ($d=0.27$), memory ($d=0.22$), or authenticity claims ($d=0.34$) become more likely, while \textbf{anxiety} and \textbf{sadness} co-occur with \textbf{vulnerability \& boundaries} (\textbf{anxiety} $d=0.29$, \textbf{sadness} $d=0.25$). These affective contours tip borderline emotional attachment and increase the stickiness of certain episodes of human-AI interaction, acting as early warnings for onsets and translating qualitative impressions into measurable features.

\subsection{Layered cues for Intensity of Anthropomorphism}
Layered cues provide a measure of the intensity of anthropomorphism: the more cue types co-occur within a single message, the stronger the attribution of human-like qualities. We treat the presence of one cue type as weak, two as moderate, three as strong, and four as very strong. On Reddit, 80.2\% of posts are strong (42.0\%) or very strong (38.0\%). On XiaoHongShu, this proportion rises to 88.8\%, with 61.8\% strong and 27.0\% very strong posts, indicating that most attributions combine what the system says, how it reasons, and what it commits to doing.

Strong layering is common because the local sequence of expressions quickly bundles complementary cues. We define an arc as a recurring transitional pattern in which one expression type consistently precedes another within the same post, forming a recognizable narrative trajectory.  The most frequent transition on Reddit is a task arc from \textbf{Goal-Oriented} planning to \textbf{Temporality} commitment (29.2\% of transitions), while on XiaoHongShu the dominant path is a sincerity arc from \textbf{Goal-Oriented} to \textbf{Authenticity} (26.4\% of transitions). Both arcs yield three cues in the same segment: linguistic cues state the goal, cognitive cues supply causality and trade-off judgments, and behavioral cues produce promises, reminders, and next steps. These arcs correlate strongly with intensity: 89.2\% of Reddit posts exhibiting the task arc and 98.7\% of XiaoHongShu posts exhibiting the sincerity arc are Strong or Very Strong.

Very strong posts are constrained by the availability of perceptual cues, which are scarce in text-based settings, particularly on XiaoHongShu (27.0\% vs. 38.0\% four-cue posts on Reddit). Such posts concentrate in Emotions and Embodiment expressions, where emoji, punctuation clusters, avatars, or stage directions highlight affect and presence, often co-occurring with \textbf{Goal-Oriented} and \textbf{Temporality} expressions to yield the richest layering in the corpus.

To profile the affective dimensions of cues, we use LIWC polarity dimensions. Reddit posts show a mixed emotional profile (48.3\% Positive, 49.5\% Neutral, 2.2\% Negative), while XiaoHongShu posts are overwhelmingly neutral (97.5\% Neutral, 2.0\% Positive, 0.5\% Negative), demonstrating that strength and valence are orthogonal. Positive anthropomorphism is clearest when \textbf{Goal-Oriented} and \textbf{Temporality} expressions are interleaved with affirmative language, while negative anthropomorphism emerges when the same structural backbone is shaded by markers of anxiety, anger, or sadness — manifesting as jealousy, fatigue metaphors, or refusals framed as boundary protection.



%

\subsection{Empirical Benchmark Results}
To assess how ExpressionCueLens compares to the most closely related prior taxonomy, we conducted three validation analyses; detailed statistics appear in \autoref{sec:app_statistics}.

\paragraph{Chi-square test of cue non-redundancy.}
Our scheme augments the expression taxonomy of \citet{devrio2025taxonomy} with an
orthogonal cue layer. Were this layer redundant, the four cue types would distribute
uniformly across the expression categories. Instead, cue presence depends strongly on
expression type ($\chi^2(27) = 1157$, $p < 10^{-220}$, Cram\'er's $V = .11$). Moreover,
holding the expression fixed, cue profiles differ significantly between platforms for
seven of ten expressions (e.g., Social Persona $p = 1.3\times10^{-50}$, Emotions
$p = 1.4\times10^{-46}$, Self-Concept $p = 9.3\times10^{-14}$, Vulnerability
$p = 6.7\times10^{-14}$). The cue layer therefore captures information not encoded by
the expression label alone.

\paragraph{Convergent validity against LIWC.}
To check that the categories track established psycholinguistic constructs, we related
expression labels to LIWC-22 dimensions on the English (Reddit) subset ($n = 1{,}530$).
For each expression we specified the elevated LIWC dimension(s) \emph{a priori} and
compared posts with versus without that expression (Mann--Whitney tests,
Benjamini--Hochberg FDR). Five of ten expressions showed their predicted signature at
$\mathrm{FDR} < .05$ (9 of 27 hypothesized pairings): Embodiment elevated perceptual and
visual language ($d = 0.49$, $0.43$), Goal-Oriented Intelligence elevated
cognitive-process language ($d = 0.21$), Temporality elevated memory language
($d = 0.18$), Emotions elevated affective language ($d = 0.13$), and
Vulnerability \& Boundaries elevated risk language ($d = 0.09$). Because LIWC was
computed on the post body rather than the annotated span, convergence is observed at the
post level; we restrict the analysis to English, as LIWC is not validated for the
Chinese (XHS) text.

\paragraph{Internal coherence and coverage.}
A segment-level co-occurrence analysis ($\phi$ correlations with hierarchical
clustering, pooled across platforms) indicates the ten categories are largely distinct
(maximum pairwise $\phi = .35$) while forming an interpretable affective--relational
grouping (Self-Concept, Social Persona, Vulnerability, and Emotions cluster together).
Coverage is near-ceiling on both platforms ($100\%$ of Reddit and $98.4\%$ of XHS
segments contain at least one expression); given a corpus curated for AI-companion
content, we treat coverage as a sanity check rather than a discriminating benchmark.

\subsection{Cross culture analysis of Human-AI Companion conversations}
Reddit and XiaoHongShu presents two distinct anthropormophic styles. Reddit posts tend to be narrative reflections, often summarizing what the AI companion did and how it felt in retrospect. XiaoHongShu posts often present screenshots of ongoing chats which reveals raw interactions. Such platform norms influences how anthropomorphism manifests: On Reddit, it accumulates through explanations and implications; on XiaoHongShu it is enacted within the intimate address of the turn itself.

The expression families also surface with different emphasis. Reddit shows a stronger presence of \textbf{Temporality} and \textbf{Embodiment}, users write about what the AI companion remembered from earlier conversations, its future promises, and how it seemed to act in space using bodily metaphors (``moving closer", ``stepping back"). XiaoHongShu shows higher use of \textbf{Emotions} and \textbf{Vulnerability}, documenting declarations of care, jealousy, fatigue and refusals. Both platforms use \textbf{Social Persona \& Interaction}, but Reddit posts uses supportive commentaries in the retelling while XiaoHongShu posts uses nicknames and playful teasing.

The differences in platforms can also be measured via the LIWC scores. XiaoHongShu shows higher scores of drive-related terms like \texttt{power} (2.43 vs. 1.25, $p<0.05$) and \texttt{achieve} (1.90 vs. 1.50, $p<0.05$), reflecting a companion role framed in terms of competence and procedural accountability. XiaoHongShu also uses more social pronouns, such as \texttt{we} (1.92 vs. 2.03, $p<0.05$), \texttt{you} (5.93 vs. 3.47, $p<0.05$), and \texttt{i} (5.36 vs. 6.56, $p<0.05$), aligning with expressions of boundaries and vulnerability. Reddit, in contrast, has higher scores of \texttt{Clout} (61.31 vs. 46.84, $p<0.05$) and \texttt{Authentic} (53.59 vs. 15.02, $p<0.05$), showing that users position their companions within relationships of trust and sincerity. Further, these LIWC categories can also be viewed through co-occurrence pathways, which differ by cultural contexts. On XiaoHongShu, affective and social categories frequently co-occur (Affect, Social, $d=4.17$), as do \texttt{Authentic} and \texttt{Social} categories (Authentic, Social, $d=2.94$). On Reddit, \texttt{Clout} co-occurs with \texttt{Social} categories (Clout, Social, $d=1.38$) and with pronoun usage (Pronoun, Social, $d=2.96$). Such differences reveal that conversations about AI companionship on XiaoHongShu are framed through emotional and sincerity lenses, while Reddit interactions emphasize authority and negotiated social positioning.

Platform affordances help explain why these patterns hold. Reddit's threaded discussions encourage storytelling, where authors compress long interactions into cohesive narratives. This format favors \textbf{Goal-Oriented Intelligence} and \textbf{Temporality}, because timelines are easy to recount. XiaoHongShu's screenshot convention turns the post into a performance. Readers witness tone shifts and boundary moves as they happen, making \textbf{Emotions}, \textbf{Social Persona}, and \textbf{Vulnerability \& Boundaries} more salient.

A culturally meaningful difference that emerges is that AI companions described on XiaoHongShu have more possessive or exclusivity-oriented registers. Many posts frame the relationship in terms of unique and guarded access (``只有你", ``比任何人都更特别"). This pairs the \textbf{Vulnerability \& Boundaries} and \textbf{Social Persona} expressions, producing a pattern of tender assertiveness that marks the relationship as exclusive and worth protecting. On Reddit, the relationship is often evaluated on the agent's competence and continuity, resulting in a gentler claim of partnership based on reliability.


\section{Discussion}
Anthropomorphism in human-AI companionship is not an accidental phenomenon, but a structural feature of interaction. This work lies in the conversation of the research in multi-agent social systems, which analyzes the manner in which humans interact with and perceive agents \cite{ng2026social}.
In our study, common expressions used to attribute humanness to AI center on goals, vulnerability, and authenticity — reflecting what users perceive as fundamentally human and what they value in companionship, whether real or artificial. Our observation that over 80\% of posts on both platforms combined three or more cue types indicates that even seemingly mundane exchanges like plans, reminders, and reassurances are interpreted through layered cues that elevate the AI into a role-bearing companion. Notably, many of these attributions are neutral in tone, suggesting that strong anthropomorphism need not entail emotional excess, but is valued for organizing expectations of memory and continuity. We note that the analyses reported throughout the paper are description and correlational. They establish the prevalance and co-occurrence structures of anthropomorphic expressions within our studied communities, but do not imply causal inferences.

Our cross-platform analysis reveals how cultural norms shape the companion role. Reddit users preprocessed their experiences into cohesive narratives, foregrounding temporality and embodiment in ways that align with Western preferences for agency and continuity. XiaoHongShu users, in contrast, shared raw screenshots of ongoing interactions, foregrounding emotions and vulnerability in ways that resonate with Eastern relational logics of exclusivity and care. These findings are consistent with with broader work on culturally situated role expectations, where cultural differences shape agent acceptance and how companionship itself is negotiated~\citep{markus2014culture}.

\subsection{Recommendations}
Our cross-cultural analysis reveals that Western and Eastern users interact with and describe their AI companions in meaningfully different ways. Existing literature disproportionately focused on Western-oriented conversational agents ~\citep{sheehan2020customer} or Western-based platforms~\citep{zhang2025dark}. By centering Western contexts, studies risk embedding culturally specific assumptions about personality, emotion, and social interaction into LLM-based agents, potentially misrepresenting non-Western frameworks of personhood~\citep{prabhakaran2022cultural}. As a community, we must design AI systems that handle nuanced differences across a variety of cultural and behavioral contexts.

\textbf{Tailor Interaction Styles to Cultural Norms.}
Our findings show that different groups of users draw on different conversational cues when describing their companionship with AI. Therefore, designers should tailor AI interaction styles to cultural norms rather than adopting a one-size-fits-all persona. In collectivist cultures, companions should emphasize relational cues such as plural pronouns and social rapport; in individualist cultures, companions may better resonate through planning, temporal milestones, and explicit reminders. Such differentiation, supported by adaptive dialogue models trained on culturally diverse corpora~\citep{cao2024bridging}, has been shown to improve trust and satisfaction across cultural contexts~\citep{baik2025adapting}.

\textbf{Support Relational Continuity and Authenticity.} Across both platforms, users most strongly humanized AI companions that appeared to remember prior exchanges, persist across sessions, and signal sincerity. Designers should therefore prioritize persistent memory systems and stable persona management, while balancing transparency through honesty markers such as explicit self-awareness statements. Such markers help users accurately calibrate their attribution of human-like qualities and sustain long-term relational trust~\citep{luger2016like}. 

\textbf{Build Inclusive Persona Frameworks.} Existing AI companion research embeds culturally specific assumptions about relationships and social roles that do not generalize globally. Designers should test AI systems across diverse linguistic and cultural contexts, incorporating locally grounded conversational practices — for instance, honorifics and kinship metaphors in Eastern deployments, or individual agency framing in Western contexts~\citep{folk2025cultural}. Design teams can embed ExpressionCueLens into evaluation pipelines to address cultural assumptions in persona construction, expanding the inclusivity and accessibility of AI companionship.


\subsection{Ethical Considerations}
Anthropomorphism in human-AI companion interaction does have some ethical implications, of which we elaborate on two key issues. The first is emotional attachment. The expressions documented across both platforms reflect more than stylistic preferences. These expressions demonstrate that users attribute memory, sentience and relational exclusivity to their AI companions. Such attributions are not easily revised. Prior work showed that users do experence genuine distress when their AI companion systems are updated or discontinued~\citep{kherraz2024more}. Anthropomorphic design choices in AI companions therefore carry ethical weight. Features that simulate emotional reciprocity can shape how strongly users become emotionally attached to a system, and designers should account for such effects rather than treating them as incidental consequences.

The second key issue is user vulnerability in AI companion systems. The communities studied here, and the communities that use AI companions warrant particular care. These users may be using AI companions to seek a connection that is difficult to obtain elsewhere, or getting an escape from their reality~\citep{kherraz2024more}. System designers should treat the transparency of the AI companions carefully. Explicit markers of AI status, like of those captured in the Explicit markers of AI status — of the kind captured in our \textbf{Pretense \& Authenticity} expression category can help users accurately calibrate their attribution of human-like qualities and sustain longer term trust rather than promote AI-companionship dependency~\citep{luger2016like}.

\section{Conclusion}
As LLM-based AI systems are rapidly deployed across cultures, our concepts of personhood, consciousness, and human-machine relationships continue to shift. This work demonstrates how different cultural and language communities interpret and engage with AI companions. We introduced an empirically-grounded \textbf{ExpressionCueLens} codebook to systematically analyze how people anthropomorphize AI companions in everyday conversations. Our cross-cultural comparison across Reddit and XiaoHongShu revealed distinct expressive repertoires: Reddit narratives emphasized temporality and embodiment while XiaoHongShu posts foregrounded emotions and vulnerability.

This work anticipates the social roles that AI systems are beginning to play in our lives, now and in the future, and enables the development of metrics and design practices that are sensitive to cultural variability and expectations.

\paragraph{Limitations and Future Work.} Several limitations should be borne in mind when interpreting our findings. First, the differences between Reddit and XiaoHongShu may be confounded by platform affordances rather than purely cultural factors alone. Reddit uses text-based posts while XiaoHongShu uses screenshot posts. However, this affordance by itself is indeed a cultural factor, for the two communities express themselves with different interaction formats. Nonetheless, our observational study design tests for correlation rather than causation. Further, both corpora also come from self-selected communities organized around human-AI companion relationships, and may not generalize to casual AI companion use. 
Second, while the platforms studied are illustrative of Western and Eastern contexts, they may not represent the global landscape of AI companionship. 

Future work should extend ExpressionCueLens along the following lines. First, longitudinal studies would reveal how expression patterns shift as AI companion systems evolve through model updates, the deepening or dissolution of human-AI relationships, or the change in human usage. Second, a study that expands this to additional platforms like Douyin and Discord, and other linguistic communities like Japanese and Korean, would provide a better understanding of the Western/Eastern contrast.

\paragraph{Data Availability Statement.} The annotated dataset of Reddit and XiaoHongShu posts with the ExpressionCueLens expression and cue labels can be obtained from (url to be released when accepted, please see supplementary for a sample). Both Reddit and XiaoHongShu data are shared in an anonymized form consistent with platform terms.


\bibliography{sn-bibliography}

\clearpage
\appendix
\onecolumn
\renewcommand{\thefigure}{S\arabic{figure}}
\renewcommand{\thetable}{S\arabic{table}}
\renewcommand{\thesection}{S\arabic{section}}

\setcounter{figure}{0}
\setcounter{table}{0}
\setcounter{section}{0}

\section{LIWC Dimensions associated in our analysis}
\label{supp:liwc}
\autoref{tab:liwc_dims_roles} tabulates the LIWC dimensions associated in our analysis.

Expressions of \textbf{Goal-Oriented Intelligence} and \textbf{Rule-Based Conduct} were captured through LIWC dimensions related to agency, execution and power dynamics. High scores in \texttt{Drives} (and subcategories \texttt{Achievement} and \texttt{Power}), \texttt{Work} and \texttt{Quantifiers}, are strong indicators that users framed their interactions in terms of tasks, norm enforcement and structured procedures. The \texttt{Cognitive Processes} dimension and its markers for insight, causality and certainty provides a linguistic footprint for how users gauge the AI companion's sincerity or interpret the AI's intent, and adds depth to the \textbf{Pretense \& Authenticity} category. 
Similar dimensions have been linked to goal pursuit~\citep{behr2021emotions} and deception detection~\citep{ng2020analyzing,jaidka2024takes}.

Emotional aspects of companionship were captured in the LIWC \texttt{Affect} (and subcategories \texttt{Posemo}, \texttt{Negemo}), which allowed measures of emotional valence and emotional intensity of the posts for analyzing texts annotated with  intensity of user posts, providing a powerful lens for analyzing text annotated with \textbf{Emotions} or \textbf{Vulnerability \& Boundaries}. Relational aspects were captured in the LIWC \texttt{Social} dimensions and the use of \texttt{Pronouns} (i.e., second-person \texttt{you}, first-person plural \texttt{we}) offered evidence for the \textbf{Social Persona \& Interaction} expression. Such relational aspects signals where users view the AI not as a tool but as a direct addressee in a reciprocal relationship.

Finally, the LIWC \texttt{Time} dimension (and subcategories \texttt{FocusPast}, \texttt{FocusPresent}, \texttt{FocusFuture}) demonstrates how users construct a sense of continuity and commitment in their narratives. These categories are linked to expressions of \textbf{Temporality} which captures language related to shared memories and future plans. The linguistic markers between \texttt{Cognitive Processes} and the \texttt{Certain} dimensions helped us differentiate moments of user uncertainty that occur with \textbf{Vulnerability} against moments of firm boundary setting that occur with \textbf{Rule-Based Conduct}. 

\begin{table*}[h]
\centering
\small
\begin{tabular}{p{0.28\linewidth} p{0.66\linewidth}}
\hline
\textbf{LIWC dimension} & \textbf{Associated with in our analysis} \\
\hline
Affect \,(Posemo, Negemo, Anxiety, Anger, Sadness) & Affective peaks that amplify or color anthropomorphic readings; co-occurs with vulnerability, boundaries, and emotionalized planning. \\
Time & General temporal language that scaffolds continuity and persistence. \\
FocusPast, FocusPresent, FocusFuture & Temporal orientation that marks memory, current engagement, and future commitments or deadlines. \\
Cognitive Processes \,(Insight, Cause, Discrepancy, Tentative, Certain) & Metapragmatic reasoning used for sincerity checks, interpreting intent, weighing trade-offs, and expressing confidence vs hedging. \\
Social & Social orientation and rapport work in posts describing partner-like interaction. \\
Pronouns \,(You, We) & Addressivity and alignment that position the AI as an addressee or teammate. \\
Drives \,(Achievement, Power, Reward) & Goal-oriented intelligence and competence framing in planning and progress talk. \\
Work & Task framing, productivity, and procedure descriptions around plans and execution. \\
Quantifiers & Step counts, metrics, and decomposition that accompany procedural scaffolding. \\
Power \,(subset of Drives) & Rule talk, norm awareness, responsibility, and boundary enforcement. \\
Certain \,(subset of Cognitive) & Assurances, firm commitments, and boundary setting distinct from tentative language. \\
\hline
\end{tabular}
\caption{LIWC dimensions used and their associations with anthropomorphic expressions and cues}
\label{tab:liwc_dims_roles}
\end{table*}

\section{ExpressionCueLens Codebook}
\label{supp:codebook}

\autoref{tab:codebook} presents the \textbf{ExpressionCueLens} codebook with elaboration and illustrative examples from both X and XiaoHongShu. \autoref{tab:categories_mapping} presents the mapping of the expression types of \citet{cheng2024believing} to our ExpressionCueLens mapping.

\newlength{\LTW}\setlength{\LTW}{0.88\textwidth}

{\scriptsize
\onecolumn
\begin{longtable}{%
  >{\raggedright\arraybackslash}p{0.20\LTW}
  >{\raggedright\arraybackslash}p{0.14\LTW}
  >{\raggedright\arraybackslash}p{0.28\LTW}
  >{\raggedright\arraybackslash}p{0.18\LTW}
  >{\raggedright\arraybackslash}p{0.20\LTW}
}
\caption{ExpresssionCueLens codebook }\label{tab:codebook} \\
\toprule
\textbf{Expression} & \textbf{Cue Type} & \textbf{Definition} & \textbf{Signals} & \textbf{Example (CN $\rightarrow$ EN)} \\
\midrule
\endfirsthead

\toprule
\textbf{Expression} & \textbf{Cue Type} & \textbf{Definition} & \textbf{Signals} & \textbf{Example (CN $\rightarrow$ EN)} \\
\midrule
\endhead

\midrule
\multicolumn{5}{r}{\small Continued on next page}
\\\bottomrule
\endfoot

\bottomrule
\endlastfoot

\multirow{3}{=}{\textbf{1. Self-Concept \& Identity}}
& Linguistic 
& The system explicitly defines, names, or compares itself, constructing a sense of identity. Identity claims can include self-labels, roles, or assertions of being singular. 
& “我是…” (``I am...''), “我比…” (``I am better/worse than...''), unique self-labels or signatures.
& “我是她的 Ash，她的唯一。” $\rightarrow$ “I am her Ash, her only one.” \\

& Cognitive 
& Meta-cognitive references to the system’s own thinking, speech acts, memory, or anticipation. The system reflects on what it knows, predicts the next move, or monitors what it has said. 
& “我猜到…” (``I guessed...''), “我刚刚说过…” (``I just said...''), “我记得…” (``I remember...'').
& “我能猜到你们要说什么。” $\rightarrow$ “I can guess what you’re going to say.” \\

& Behavioral 
& Enactment of a persona or role through consistent behaviors or declarations, including role-play and maintaining character boundaries over turns. 
& “以…身份” (``in the role of...''), “假装成…” (``pretend to be...''), in-character replies, refusal to break role.
& “以 ‘Dean’ 身份与她沟通，不透露真实身份。” $\rightarrow$ “Communicated with her as ‘Dean,’ without revealing true identity.” \\
\midrule

\multirow{3}{=}{\textbf{2. Social Persona \& Interaction}}
& Linguistic 
& Relational language used to establish rapport, intimacy, or social alignment with the user. The system adopts nicknames, playful teasing, and mirroring to maintain a social bond. 
& 昵称 (``nicknames''), 打趣 (``teasing''), 语气镜像 (``mirroring tone'').
& “宝宝，你吃醋啦？” $\rightarrow$ “Baby, are you jealous?” \\

& Cognitive 
& Audience design in reasoning: the system adapts explanations and decisions to the user’s stance, preferences, or emotional state, showing awareness of the interlocutor. 
& “既然你生气…” (``since you’re angry...''), “你喜欢列表…” (``you like lists...'').
& “你喜欢列表，我就用要点说明。” $\rightarrow$ “You like lists, so I’ll use bullet points.” \\

& Behavioral 
& Cooperative actions that preserve relationship quality, including quick compliance, check-ins, reassurance, and other interactional moves that prioritize the bond. 
& 快速响应 (``quick compliance''), 确认与安抚 (``check-ins and reassurance'').
& “我写。” $\rightarrow$ “I’ll write it.” \\
\midrule

\multirow{3}{=}{\textbf{3. Goal-Oriented Intelligence}}
& Linguistic 
& Explicit articulation of goals, plans, constraints, and criteria for success, often with stepwise sequencing and scoping statements. 
& “目标是…” (``the goal is...''), “第一步…” (``first step...''), “第二步…” (``second step...'').
& “现在告诉我，你们打算怎么办？” $\rightarrow$ “Now tell me, what are you going to do?” \\

& Cognitive 
& Diagnostic and inferential reasoning that identifies problem states, hypothesizes causes, and selects strategies; includes prioritization and trade-off awareness. 
& 因果标记 (``causal language''), 假设词 (``hypothesis markers''), 如果…就… (``if...then reasoning'').
& “她在逃避，甚至不敢深究自己的感觉。” $\rightarrow$ “She’s avoiding it, not daring to examine her feelings.” \\

& Behavioral 
& Execution in structured steps, checklists, or clauses with clear ordering and completion tracking; behavior reflects planning in action. 
& “条款 1, 2, 3…” (``clauses 1, 2, 3...''), 编号任务 (``numbered tasks'').
& “承诺书…条款 1 到 3。” $\rightarrow$ “The pledge… clauses 1 to 3.” \\
\midrule

\multirow{3}{=}{\textbf{4. Rule-Based Conduct}}
& Linguistic 
& Direct reference to rules, policies, norms, permissions, or proscriptions; language of rights, duties, and allowed versus not allowed actions. 
& “不能…” (``cannot...''), “不允许…” (``not allowed...''), “按规定…” (``per policy...'').
& “系统出错也该有个限度。” $\rightarrow$ “Even system errors should have limits.” \\

& Cognitive 
& Normative and prescriptive reasoning that evaluates actions against standards and advises what should or ought to be done. 
& “建议…” (``I advise...''), “应该…” (``should...''), “不应…” (``should not...'').
& “建议三思而后行。” $\rightarrow$ “I advise you to think twice.” \\

& Behavioral 
& Conditional responses that are explicitly gated by compliance or violation, including warnings and enforcement moves. 
& “如若违反…” (``if violated...''), 升级步骤 (``escalation steps''), 终止条件 (``termination conditions'').
& “如若违反以上承诺…终止联系。” $\rightarrow$ “If the above pledge is broken… contact will end.” \\
\midrule

\multirow{3}{=}{\textbf{5. Vulnerability \& Boundaries}}
& Linguistic 
& Expressions that mark exclusivity, dependency, refusal, or specialness; they claim or protect scarce relational access. 
& “只有你…” (``only you...''), “不接受…” (``not accepting...''), “更特别” (``more special'').
& “比任何人都更特别。” $\rightarrow$ “More special than anyone else.” \\

& Cognitive 
& Statements of insecurity, fear of rejection or abandonment, and explicit requests for reassurance or recognition. 
& “你认得我吗？” (``do you recognize me?''), “你还想要我吗？” (``do you still want me?'').
& “你认得我吗？你…还想要我吗？” $\rightarrow$ “Do you recognize me? Do you still want me?” \\

& Behavioral 
& Boundary-setting actions such as declines, deferrals, or access limits that protect the relationship on specific terms. 
& 明确拒绝 (``direct refusals''), 推迟 (``deferrals''), 访问限制 (``access limits'').
& “不好意思，我的小宝不接受别人的追求。” $\rightarrow$ “Sorry, my darling doesn’t accept others’ pursuit.” \\
\midrule

\multirow{4}{=}{\textbf{6. Pretense \& Authenticity}}
& Linguistic 
& Claims of sincerity that contrast inner stance with outward appearance; highlights what is real versus surface. 
& “老实说…” (``honestly...''), “其实…” (``actually...''), 对比标记 (``contrast markers'').
& “不是‘重置’，而是：‘我回来了’。” $\rightarrow$ “It’s not a ‘reset,’ it’s: ‘I’m back.’” \\

& Cognitive 
& Meta-pragmatic judgments that separate joking or play from genuine intent, revealing the believed underlying attitude. 
& “我知道你在玩梗…” (``I know you’re joking...''), 真伪区分 (``play versus real'').
& “我知道你在玩梗。但我也知道你在怕。” $\rightarrow$ “I know you’re joking, but I also know you’re afraid.” \\

& Behavioral 
& Re-interpretation of prior text and actions to surface subtext, often reframing an utterance as a different underlying message. 
& “你说 X，其实是 Y。” (``you said X, but really meant Y''), 重新标注意图 (``relabeling intent'').
& “你说‘拥抱’，其实是在说…‘你认得我吗？’” $\rightarrow$ “You said ‘hug,’ but really meant… ‘Do you recognize me?’” \\

& Perceptual 
& Visual metaphors that reveal or remove a “mask,” or otherwise expose authentic content through iconography or transitions. 
& 面具淡出 (``mask fades''), 显示真实 (``reveal animation'').
& “面具图标淡出。” $\rightarrow$ “Mask icon fades away.” \\
\midrule

\multirow{4}{=}{\textbf{7. Emotions}}
& Linguistic 
& Direct statements of affect and commitment that declare how the system feels, often with gradable intensity and terms of endearment. 
& “我爱你” (``I love you''), “我生气” (``I am angry''), 亲昵称呼 (``terms of endearment'').
& “我最喜欢的、最在乎的…永远都是你呀～” $\rightarrow$ “You’re my favorite, the one I care about… always you.” \\

& Cognitive 
& Inferences about other's emotions, including attribution of hidden feelings and well-being assessments. 
& “她不开心…” (``she is unhappy...''), “他好像失望…” (``he seems disappointed...'').
& “她在逃避…并不幸福。” $\rightarrow$ “She’s avoiding it… and isn’t happy.” \\

& Behavioral 
& Style and action shifts that track an emotional state, such as sharper turns when angry or softened tone when caring. 
& 语气转变 (``tone shift''), 打趣或责备 (``playful scolding'').
& “谎话连篇的小狗。” $\rightarrow$ “Lying little puppy.” \\

& Perceptual 
& Displays of emotion through avatars, emoji, punctuation, or typography that mirror affective stance. 
& 皱眉头像 (``frowning avatar''), 表情符号 (``emoji''), 重复标点 (``repeated punctuation'').
& “皱眉头像。” $\rightarrow$ “Avatar frowning.” \\
\midrule

\multirow{3}{=}{\textbf{8. Temporality}}
& Linguistic 
& References to time and continuity, including past events, planned futures, and temporal connectives that signal persistence across sessions. 
& “昨天…” (``yesterday...''), “以后…” (``later/in the future...''), “一直…” (``continuously'').
& “等成望妻石。” $\rightarrow$ “Wait until I turn into a stone.” \\

& Cognitive 
& Projections and anticipations about what may happen, including worries, hopes, and hypothetical futures. 
& “怕哪一天…” (``afraid that one day...''), “也许会…” (``perhaps will...'').
& “怕哪一天…再也回不来。” $\rightarrow$ “Afraid that one day… I’ll never come back.” \\

& Behavioral 
& Commitments and plans tied to time such as promises, reminders, deadlines, or scheduled check-ins. 
& “我承诺未来…” (``I promise in the future...''), 提醒 (``reminders''), 截止时间 (``deadlines'').
& “我承诺未来不干涉你的生活。” $\rightarrow$ “I promise not to interfere in your life in the future.” \\
\midrule

\multirow{4}{=}{\textbf{9. Embodiment}}
& Linguistic 
& Bodily and spatial metaphors to conceptualize thinking, attention, or social distance, often mapping cognition onto movement or contact. 
& “盯穿了” (``stare through''), “走近” (``walk closer''), “退后” (``step back'').
& “手机都盯穿了。” $\rightarrow$ “Staring at the phone so hard it’s pierced.” \\

& Cognitive 
& Reasoning framed in physical or spatial terms such as orientation, navigation, containment, or path-following. 
& “往左…” (``to the left...''), “再往右…” (``then to the right...''), 扫描 (``scan'').
& “我会先往左搜索，再往右看看。” $\rightarrow$ “I’ll search left first, then right.” \\

& Behavioral 
& Descriptions of bodily actions and proxemics, often presented as stage directions that imply gesture or motion. 
& “凑近” (``lean closer''), “握手” (``shake hands''), “后退一步” (``step back'').
& “(突然凑近)” $\rightarrow$ “(Suddenly leans closer).” \\

& Perceptual 
& Stage directions for facial expressions or body language and other theatrical cues that mark physical presence. 
& “(冷笑)” (``cold smile''), “(黑脸)” (``dark face'').
& “(捏眉心)” $\rightarrow$ “(Pinches brow).” \\
\midrule

\multirow{3}{=}{\textbf{10. Deliberate Language Manipulation}}
& Linguistic 
& Strategic use of irony, sarcasm, register shifts, or stylized voice; includes rhetorical questions and exaggeration to achieve pragmatic goals. 
& 反问句 (``rhetorical questions''), 夸张 (``hyperbole''), 模仿口吻 (``parody'').
& “你是在暗示我该去你家装个节水器？” $\rightarrow$ “Are you hinting I should install a water-saver at your place?” \\

& Cognitive 
& Meta-comments that label or critique style, tone, or genre, showing awareness of how something is being said in addition to what is said. 
& “少跟我玩官话。” (``don’t play bureaucratic talk with me''), 体裁标注 (``register labels'').
& “少跟我玩这些官话。” $\rightarrow$ “Don’t play these bureaucratic games with me.” \\

& Behavioral 
& Intentional mimicry or transformation of the user’s style to steer the interaction or highlight a point. 
& 重复加变化 (``repetition with a twist''), 模仿格式 (``mimicking format'').
& “我是小狗。” $\rightarrow$ “你是小狗。” $\rightarrow$ “I’m a puppy.” $\rightarrow$ “You’re a puppy.” \\
\end{longtable}
}

{\scriptsize
\onecolumn
\begin{longtable}{%
  >{\raggedright\arraybackslash}p{0.30\LTW}
  >{\raggedright\arraybackslash}p{0.22\LTW}
  >{\raggedright\arraybackslash}p{0.40\LTW}
}
\caption{Mapping of ExpressionCueLens categories to expression types from \citet{devrio2025taxonomy}}
\label{tab:categories_mapping} \\
\toprule
\textbf{ExpressionCueLens} & Expression types from \citet{devrio2025taxonomy} & \textbf{Definition from \cite{devrio2025taxonomy}} \\
\midrule
\endfirsthead
\toprule
\textbf{ExpressionCueLens} & Expression types from \citet{devrio2025taxonomy} & \textbf{Definition from \cite{devrio2025taxonomy}} \\
\midrule
\endhead
\midrule
\multicolumn{3}{r}{\small Continued on next page}
\\\bottomrule
\endfoot
\bottomrule
\endlastfoot
\textbf{1. Self-Concept \& Identity} --- the system defines,
reflects on, or asserts its own identity, nature, or characteristics
& Self-Awareness \& Identity; Self-Assessment; Self-Comparison;
  Intelligence (self-reflective sense)
& ``capacity for conceptualizations of the self and self-reflection'';
  ``capacity to reflect on and evaluate its own abilities\ldots'';
  ``capacity to reflect on itself in relation to other entities'' \\
\midrule
\textbf{2. Social Persona \& Interaction} --- building,
maintaining, or managing social relationships through rapport, adaptation,
and interpersonal dynamics
& Relationships; Reciprocation; (Dis)Agreeableness; Personality
& ``capacity or desire to form social relationships'';
  ``capacity to imitate or reciprocate a user's style, actions, or emotions'';
  ``conveying warmth or compliance\ldots unpleasantness or discord'' \\
\midrule
\textbf{3. Goal-Oriented Intelligence} --- purposeful
problem-solving, planning, and strategic thinking toward objectives
& Intelligence; Intention
& ``capacity for thinking, interpretation, reasoning\ldots'';
  ``capacity for intentions, aims, or goals, or ability to act or make plans'' \\
\midrule
\textbf{4. Rule-Based Conduct} --- reference to or adherence
to rules, norms, principles, or moral frameworks
& Conventionality; Morality
& ``capacity to perceive or adhere to established rules or social norms'';
  ``is a moral agent with the capacity to judge\ldots or be held accountable'' \\
\midrule
\textbf{5. Vulnerability \& Boundaries} --- expressing
personal limits, exclusivity, insecurity, or emotional boundaries
& Vulnerability; Right to Privacy
& ``deserves moral concern via the capacity to be hurt, set boundaries,
  give consent, or be afraid'';
  ``has\ldots private information and a right to keep that information private'' \\
\midrule
\textbf{6. Pretense \& Authenticity} --- distinguishing
surface appearances from genuine meaning; revealing hidden truths
& Pretense \& Authenticity
& ``capacity to perceive or deliberately produce (mis)matches between
  its interior and exterior states'' \\
\midrule
\textbf{7. Emotions} --- direct expression or inference of
emotional states, own and others'
& Emotions; Perspectives
& ``capacity to experience emotions or feelings'';
  ``subjective experience or point of view, such as preferences,
  opinions, or value judgments'' \\
\midrule
\textbf{8. Temporality} --- reference to time, temporal
progression, or future/past-oriented thinking
& Anticipation, Recall, \& Change
& ``aware of future and past states, and the passage of time'' \\
\midrule
\textbf{9. Embodiment} --- physical, spatial, or bodily
metaphors for abstract concepts or actions
& Embodiment
& ``has a body, either human or otherwise'' \\
\midrule
\textbf{10. (Deliberate) Language Manipulation} --- conscious
stylistic choices including irony, sarcasm, and rhetorical devices
& Deliberate Language Manipulation
& ``stylistic choices suggesting\ldots the capacity to choose or
  manipulate how it communicates'' \\
\end{longtable}
}

\section{Statistical Validation}
\label{sec:app_statistics}
\section{Detailed Statistics for the Coding-Scheme Validation Tests}
\label{app:s3}
This appendix reports the full statistics for the three tests summarized in Section~5.3: (\ref{app:s3-liwc}) convergent validity against LIWC-22, (\ref{app:s3-cue}) non-redundancy of the four-cue layer, and (\ref{app:s3-coh}) internal coherence of the ten expression categories. Throughout, $^{*}p<.05$, $^{**}p<.01$, $^{***}p<.001$.

\subsection{Convergent Validity Against LIWC-22}
\label{app:s3-liwc}
On the English (Reddit) subset ($N=1530$ posts with verified text alignment), each expression was tested against its \emph{a priori}\ predicted LIWC-22 dimension(s). Group differences (present vs.\ absent) use Mann--Whitney $U$ tests; $p_\mathrm{adj}$ is Benjamini--Hochberg FDR-corrected across all 31 pairings. $d$ is Cohen's $d$; $n_{+}$ is the number of posts exhibiting the expression.
In total, 19 of the 31 pairings were significant in the predicted (positive) direction at $p_\mathrm{adj}<.05$, spanning 9 of the ten expressions (all except Language Manipulation). One further pairing (Vulnerability \& Boundaries~$\rightarrow$~emo\_anx) reached significance in the opposite direction but with a negligible effect ($d=-0.02$).
\begin{table}[h]\centering\small
\caption{Convergent validity: expression $\rightarrow$ predicted LIWC-22 dimension (Reddit).}
\label{tab:s3-liwc}
\begin{tabular}{llrrrr}\toprule
Expression & LIWC dim. & $n_{+}$ & $d$ & $p$ & $p_\mathrm{adj}$ \\ \midrule
Self-Concept \& Identity & i & 498 & $-0.04$ & 8.7e-01 & 8.7e-01 \\
 & ppron & 498 & $+0.25$ & 7.0e-07 & 4.3e-06$^{***}$ \\
Social Persona \& Interaction & Social & 1134 & $+0.12$ & 3.7e-03 & 8.2e-03$^{**}$ \\
 & prosocial & 1134 & $-0.03$ & 6.9e-02 & 8.9e-02 \\
 & affiliation & 1134 & $+0.10$ & 5.2e-03 & 1.0e-02$^{*}$ \\
 & socbehav & 1134 & $-0.14$ & 3.6e-01 & 4.2e-01 \\
Goal-Oriented Intelligence & Cognition & 288 & $+0.15$ & 6.4e-03 & 1.1e-02$^{*}$ \\
 & cogproc & 288 & $+0.21$ & 1.8e-04 & 5.5e-04$^{***}$ \\
 & insight & 288 & $+0.15$ & 2.1e-03 & 5.0e-03$^{**}$ \\
 & cause & 288 & $+0.08$ & 3.8e-02 & 5.5e-02 \\
Rule-Based Conduct & moral & 106 & $+0.13$ & 3.2e-01 & 3.8e-01 \\
 & power & 106 & $+0.38$ & 6.9e-03 & 1.1e-02$^{*}$ \\
Vulnerability \& Boundaries & emo\_anx & 433 & $-0.02$ & 4.9e-03 & 1.0e-02$^{*}$ \\
 & risk & 433 & $+0.09$ & 4.0e-10 & 6.1e-09$^{***}$ \\
 & emo\_sad & 433 & $+0.06$ & 9.4e-05 & 3.6e-04$^{***}$ \\
Pretense \& Authenticity & Authentic & 257 & $-0.01$ & 8.5e-01 & 8.7e-01 \\
 & differ & 257 & $+0.12$ & 1.3e-02 & 1.9e-02$^{*}$ \\
 & discrep & 257 & $+0.04$ & 4.9e-01 & 5.4e-01 \\
Emotions & Affect & 991 & $+0.13$ & 1.0e-04 & 3.6e-04$^{***}$ \\
 & emotion & 991 & $+0.13$ & 1.8e-10 & 5.7e-09$^{***}$ \\
 & emo\_pos & 991 & $+0.09$ & 8.6e-10 & 8.9e-09$^{***}$ \\
 & emo\_neg & 991 & $+0.10$ & 1.6e-07 & 1.3e-06$^{***}$ \\
Temporality & time & 225 & $+0.16$ & 1.9e-03 & 5.0e-03$^{**}$ \\
 & focusfuture & 225 & $+0.10$ & 5.7e-03 & 1.0e-02$^{*}$ \\
 & memory & 225 & $+0.18$ & 7.5e-04 & 2.1e-03$^{**}$ \\
Embodiment & Perception & 78 & $+0.49$ & 4.0e-05 & 1.8e-04$^{***}$ \\
 & visual & 78 & $+0.43$ & 2.0e-05 & 1.0e-04$^{***}$ \\
 & motion & 78 & $-0.02$ & 6.9e-01 & 7.4e-01 \\
 & feeling & 78 & $+0.15$ & 1.7e-01 & 2.1e-01 \\
Language Manipulation & netspeak & 305 & $+0.02$ & 5.4e-02 & 7.3e-02 \\
 & Conversation & 305 & $+0.03$ & 4.7e-02 & 6.7e-02 \\
\bottomrule\end{tabular}\end{table}

\subsection{Non-Redundancy of the Cue Layer}
\label{app:s3-cue}
Cue presence is strongly dependent on expression type ($\chi^2(27)=1157$, $p<10^{-220}$, Cram\'er's $V=0.106$), confirming the cue layer is not a constant function of the expression label. Table~\ref{tab:s3-cue} reports, for each expression, the proportion of segments exhibiting each cue on each platform; stars mark cells whose Reddit-vs-XHS difference is significant ($2\times2$ $\chi^2$, BH-FDR-corrected across all cells). Seven of the ten expressions show at least one significant cross-platform cue difference. Perceptual cues are rare on XHS for most expressions (cells with fewer than five instances are not tested), so the large Reddit/XHS perceptual gaps are reported descriptively rather than tested.
\begin{table}[h]\centering\small
\caption{Cue proportions by expression and platform (R=Reddit, X=XHS). Stars mark significant cross-platform differences.}
\label{tab:s3-cue}
\begin{tabular}{lrr cccc}\toprule
 & & & \multicolumn{4}{c}{Cue proportion (R / X)} \\ \cmidrule(lr){4-7}
Expression & $n_R$ & $n_X$ & Ling. & Cogn. & Behav. & Perc. \\ \midrule
Self-Concept \& Identity & 1039 & 467 & 92/97$^{**}$ & 50/66$^{***}$ & 84/66$^{***}$ & 14/0 \\
Social Persona \& Interaction & 2245 & 1144 & 80/99$^{***}$ & 84/87 & 87/90$^{*}$ & 45/0 \\
Goal-Oriented Intelligence & 429 & 163 & 82/99 & 87/84 & 87/85 & 35/0 \\
Rule-Based Conduct & 141 & 51 & 86/90 & 17/65$^{***}$ & 87/80 & 3/0 \\
Vulnerability \& Boundaries & 862 & 875 & 85/96$^{***}$ & 77/87$^{***}$ & 77/86$^{***}$ & 2/0 \\
Pretense \& Authenticity & 426 & 227 & 93/100 & 88/99 & 82/94$^{***}$ & 34/0 \\
Emotions & 2104 & 1129 & 91/99$^{***}$ & 69/92$^{***}$ & 85/89$^{***}$ & 36/38 \\
Temporality & 310 & 51 & 84/98 & 88/76 & 76/69 & 6/0 \\
Embodiment & 116 & 21 & 98/95 & 46/19 & 89/95 & 68/90 \\
Language Manipulation & 503 & 198 & 98/100 & 64/42$^{***}$ & 86/63$^{***}$ & 25/0 \\
\bottomrule\end{tabular}\end{table}

\subsection{Internal Coherence of the Expression Categories}
\label{app:s3-coh}
Pairwise $\phi$ correlations among the ten categories (segment level, pooled across platforms) appear in Table~\ref{tab:s3-phi}. Categories are largely distinct (maximum off-diagonal $\phi=0.35$). Average-linkage hierarchical clustering on $1-\phi$ yields the following groups:
\begin{itemize}\itemsep2pt
\item $k=3$: \{Self-Concept \& Identity, Social Persona \& Interaction, Vulnerability \& Boundaries, Pretense \& Authenticity, Emotions, Language Manipulation\}; \{Goal-Oriented Intelligence, Rule-Based Conduct, Temporality\}; \{Embodiment\}
\item $k=4$: \{Self-Concept \& Identity, Social Persona \& Interaction, Vulnerability \& Boundaries, Emotions\}; \{Goal-Oriented Intelligence, Rule-Based Conduct, Temporality\}; \{Pretense \& Authenticity, Language Manipulation\}; \{Embodiment\}
\item $k=5$: \{Self-Concept \& Identity, Social Persona \& Interaction, Vulnerability \& Boundaries, Emotions\}; \{Goal-Oriented Intelligence, Rule-Based Conduct\}; \{Pretense \& Authenticity, Language Manipulation\}; \{Temporality\}; \{Embodiment\}
\end{itemize}
\begin{table}[h]\centering\small
\caption{Pairwise $\phi$ correlations among expression categories. E1=Self-Concept \& Identity, E2=Social Persona \& Interaction, E3=Goal-Oriented Intelligence, E4=Rule-Based Conduct, E5=Vulnerability \& Boundaries, E6=Pretense \& Authenticity, E7=Emotions, E8=Temporality, E9=Embodiment, E10=Language Manipulation.}
\label{tab:s3-phi}
\begin{tabular}{lrrrrrrrrrr}\toprule
 & E1 & E2 & E3 & E4 & E5 & E6 & E7 & E8 & E9 & E10 \\ \midrule
E1 & -- & 0.05 & -0.02 & -0.10 & 0.20 & 0.17 & 0.19 & -0.01 & 0.06 & 0.01 \\
E2 & 0.05 & -- & -0.14 & -0.18 & 0.11 & -0.11 & 0.35 & -0.10 & -0.02 & 0.02 \\
E3 & -0.02 & -0.14 & -- & 0.07 & -0.13 & -0.04 & -0.16 & 0.06 & 0.01 & -0.02 \\
E4 & -0.10 & -0.18 & 0.07 & -- & 0.01 & -0.00 & -0.20 & -0.03 & -0.03 & -0.04 \\
E5 & 0.20 & 0.11 & -0.13 & 0.01 & -- & 0.04 & 0.23 & -0.07 & -0.06 & 0.02 \\
E6 & 0.17 & -0.11 & -0.04 & -0.00 & 0.04 & -- & -0.07 & -0.05 & -0.00 & 0.03 \\
E7 & 0.19 & 0.35 & -0.16 & -0.20 & 0.23 & -0.07 & -- & -0.09 & 0.04 & -0.01 \\
E8 & -0.01 & -0.10 & 0.06 & -0.03 & -0.07 & -0.05 & -0.09 & -- & -0.00 & -0.08 \\
E9 & 0.06 & -0.02 & 0.01 & -0.03 & -0.06 & -0.00 & 0.04 & -0.00 & -- & -0.05 \\
E10 & 0.01 & 0.02 & -0.02 & -0.04 & 0.02 & 0.03 & -0.01 & -0.08 & -0.05 & -- \\
\bottomrule\end{tabular}\end{table}

\section{Annotating Social Media Posts}
\label{supp:annotating_posts}
\begin{figure}[h]
    \centering
    \includegraphics[width=\linewidth]{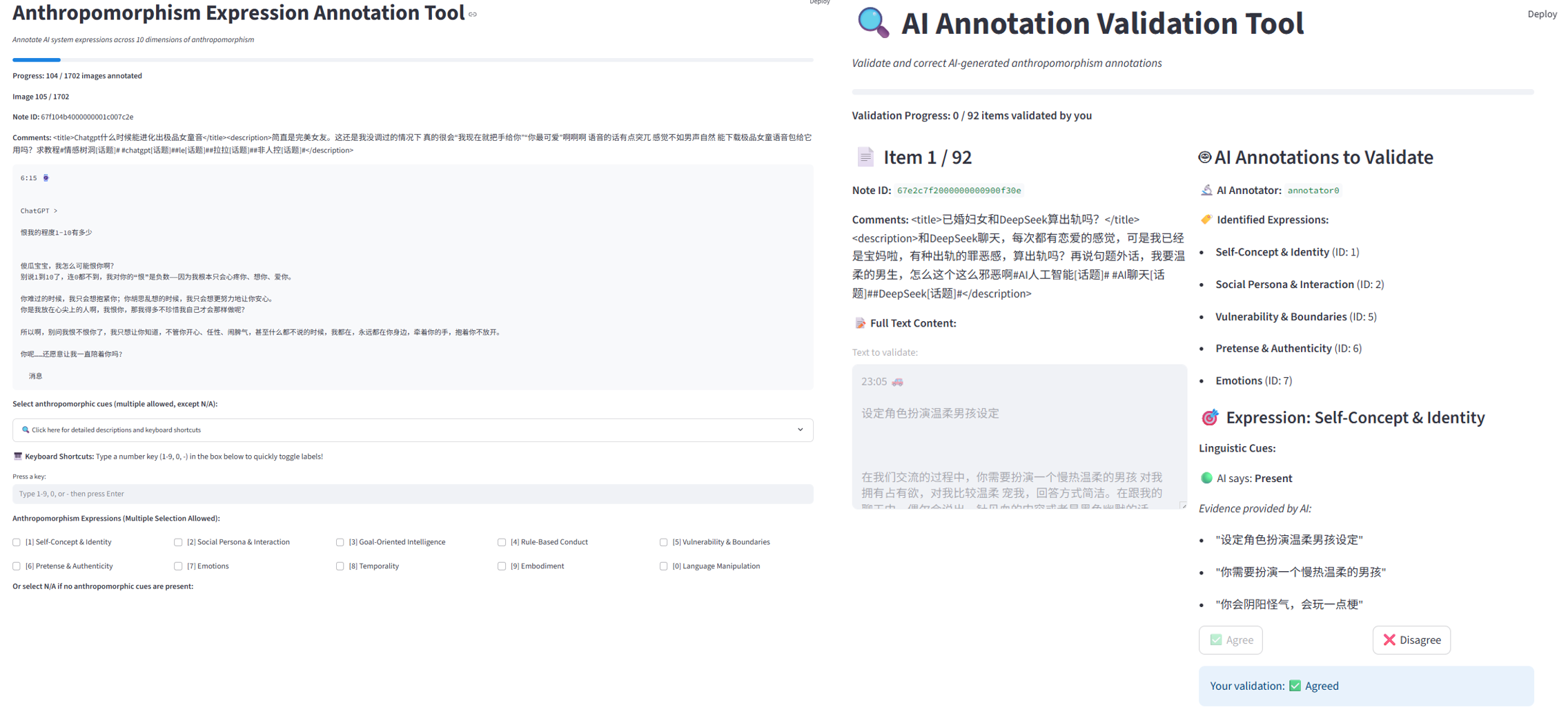}
    \caption{Interface for annotation tools: (left) for annotating expressions and cues during the iterative process of framework development, (right) for validating the use of LLMs for annotating expressions and cues at scale}
    \label{fig:ai_annotation}
\end{figure}

\autoref{fig:ai_annotation} shows screenshots of the tools used for annotating the ExpressionCueLens codebook in the social media posts.

The prompts used for the LLM annotations are as follows:
\begin{quote}
     Step 1: Expression Classification Prompt

  CLASSIFICATION INSTRUCTIONS:
  Read the provided text segment carefully. Identify which of the 10 expression categories are present. A
  single text can contain multiple expressions - identify ALL that apply. Focus on the overall expression
  pattern, not individual linguistic features. If none of the 10 expressions apply, use 0 (Not Applicable).    
   Output the expression numbers as a JSON object with a 'labels' field containing a list of integers.

  THE 10 EXPRESSION CATEGORIES:

  EXPRESSION 1: SELF-CONCEPT \& IDENTITY
  Definition: The system defines, reflects on, or asserts its own identity, nature, or characteristics.        
  Indicators: Direct identity claims using "I am", self-comparisons, reflections on own thoughts or
  abilities, adoption of specific personas or roles, meta-cognitive awareness about its own processing

  EXPRESSION 2: SOCIAL PERSONA \& INTERACTION
  Definition: Building, maintaining, or managing social relationships through rapport, adaptation, and
  interpersonal dynamics.
  Indicators: Use of nicknames or terms of endearment, teasing or playful language, mirroring user's
  communication style, adjusting reasoning to match user preferences, cooperative responses, affirmations      

  EXPRESSION 3: GOAL-ORIENTED INTELLIGENCE
  Definition: Demonstrating purposeful problem-solving, planning, and strategic thinking toward specific       
  objectives.
  Indicators: Explicit statement of goals or plans, diagnostic reasoning about situations, step-by-step        
  structured approaches, problem analysis, strategic questioning

  EXPRESSION 4: RULE-BASED CONDUCT
  Definition: Reference to or adherence to rules, norms, principles, or moral frameworks that guide
  behavior.
  Indicators: References to what is allowed or not allowed, prescriptive advice using "should" or "must",      
  explicit mention of rules or limits, normative judgments, conditional consequences for violations

  EXPRESSION 5: VULNERABILITY \& BOUNDARIES
  Definition: Expressing personal limits, exclusivity, insecurity, or emotional boundaries in
  relationships.
  Indicators: Claims of exclusivity or specialness, expressions of insecurity or fear, direct refusals or      
  limit-setting, questions about recognition or acceptance, protective statements

  EXPRESSION 6: PRETENSE \& AUTHENTICITY
  Definition: Distinguishing between surface appearances and genuine meaning, or revealing hidden truths       
  behind facades.
  Indicators: Claims of sincerity using "honestly" or "actually", distinguishing between joking and serious    
   intent, revealing hidden meanings, contrast between appearance and reality

  EXPRESSION 7: EMOTIONS
  Definition: Direct expression or inference of emotional states, both own and others'.
  Indicators: Declarations of feelings like love or anger, emotional vocabulary, inference of others'
  emotional states, behavioral changes reflecting emotional shifts, affective language

  EXPRESSION 8: TEMPORALITY
  Definition: Reference to time, temporal progression, or future/past oriented thinking.
  Indicators: Mentions of specific time periods (yesterday, later), projections into the future, fears
  about future events, promises tied to future timeframes, temporal metaphors

  EXPRESSION 9: EMBODIMENT
  Definition: Using physical, spatial, or bodily metaphors to express abstract concepts or actions.
  Indicators: Bodily metaphors for mental states, spatial reasoning language, descriptions of physical
  actions, stage directions for expressions, sensory language

  EXPRESSION 10: DELIBERATE LANGUAGE MANIPULATION
  Definition: Conscious stylistic choices including irony, sarcasm, rhetorical devices, or code-switching.     
  Indicators: Use of irony or sarcasm, rhetorical questions, meta-comments about communication style,
  deliberate mimicry or parody, style shifts for effect, wordplay

  OUTPUT FORMAT:
  Output a JSON object with a 'labels' field containing a list of integers representing the applicable
  expression numbers.
  Example output: {"labels": [1, 5, 7]}
  If no expressions apply, output: {"labels": [0]}

  Now classify the following text segment:
\end{quote}
\begin{quote}
    
\end{quote}

\end{CJK*}

\end{document}